\documentclass[aps,twocolumn,floatfix,superscriptaddress,amssymb]{revtex4-1}
\usepackage{amsmath}
\usepackage{amssymb}
\usepackage{graphicx}
\usepackage{color}
\usepackage{soul}
\usepackage[utf8]{inputenc}
\usepackage[T1]{fontenc}
\usepackage{array}
\usepackage{comment}
\usepackage{textcomp,mathcomp}
\usepackage{amsmath}
\usepackage{mathtools}
 
\usepackage{braket}
\usepackage{siunitx}
\usepackage[normalem]{ulem}

\usepackage[driverfallback=dvipdfmx,colorlinks,linkcolor=blue,citecolor=blue,urlcolor=blue]{hyperref}
\usepackage{threeparttable} 

\begin{document}
	
	\title{Apparatus for producing single strontium atoms in an optical tweezer array}
	
	\author{Kai Wen}
	\thanks{These authors contributed equally.}
	\affiliation{Microelectronics Thrust, The Hong Kong University of Science and Technology (Guangzhou), Guangzhou, China} 
	
	\author{Huijin Chen}
	\thanks{These authors contributed equally.}
	\affiliation{Microelectronics Thrust, The Hong Kong University of Science and Technology (Guangzhou), Guangzhou, China} 
	
	\author{Xu Yan}
	\affiliation{Department of Physics, The Hong Kong University of Science and Technology, Clear Water Bay, Kowloon, Hong Kong, China} 
	
	\author{Zejian Ren}
	\affiliation{Microelectronics Thrust, The Hong Kong University of Science and Technology (Guangzhou), Guangzhou, China} 
	
	\author{Chengdong He}
	\affiliation{Department of Physics, The Hong Kong University of Science and Technology, Clear Water Bay, Kowloon, Hong Kong, China}
	
	\author{Elnur Hajiyev}
	\affiliation{Department of Physics, The Hong Kong University of Science and Technology, Clear Water Bay, Kowloon, Hong Kong, China}
	
	\author{Preston Tsz Fung Wong}
	\affiliation{Department of Physics, The Hong Kong University of Science and Technology, Clear Water Bay, Kowloon, Hong Kong, China}
	
	\author{Gyu-Boong Jo}
	\email[Corresponding author email: ]{gbjo@ust.hk}
	\affiliation{Microelectronics Thrust, The Hong Kong University of Science and Technology (Guangzhou), Guangzhou, China}
	\affiliation{Department of Physics, The Hong Kong University of Science and Technology, Clear Water Bay, Kowloon, Hong Kong, China}

	\date{\today}
	
	\begin{abstract} 
		
		We outline an experimental setup for efficiently preparing a tweezer array of $^{88}$Sr atoms. Our setup uses permanent magnets to maintain a steady-state two-dimensional magneto-optical trap (MOT) which results in a loading rate of up to $10^{8}$ s$^{-1}$ at 5 mK for the three-dimensional blue MOT. This enables us to trap $2\times10^{6}$ $^{88}$Sr atoms at 2 $\mu$K in a narrow-line red MOT with the $^{1}$S$_{0}$ $\rightarrow$ $^{3}$P$_{1}$ intercombination transition at 689 nm. With the Sisyphus cooling and pairwise loss processes, single atoms are trapped and imaged in 813 nm optical tweezers, exhibiting a lifetime of 2.5 minutes. We further investigate the survival fraction of a single atom in the tweezers and characterize the optical tweezer array using a release and recapture technique. Our platform paves the way for potential applications in atomic clocks, precision measurements, and quantum simulations.
		
	\end{abstract}

	\flushbottom
	
	\maketitle
	\section{Introduction}
	
	The two-electron atoms like ytterbium and strontium in a programmed optical tweezer array~\cite{Barredo.2016,Endres.2016,Kim.2016} have shown great promise in quantum metrology, simulation and computing~\cite{cooper2018alkaline,wilson2022trapping,madjarov2020high, norcia2019seconds}, as well as precision measurement in optical lattice clock~\cite{bloom2014optical}. Thanks to the rich electronic structures associated with two valence electrons, unprecedented quantum control relevant to Rydberg excitations has been demonstrated in those systems. Recently, Rydberg atoms in reconfigurable optical arrays have been demonstrated high single- and two-qubit operation fidelity ~\cite{madjarov2020high}, which opens a promising future in quantum simulation and computing based on neutral atoms~\cite{levine2018high,browaeys2020many,levine2018high,ebadi2022quantum,chen2023continuous}.
	

	In this work, we illustrate a simple versatile platform for generating cold $^{88}$Sr samples in a two-dimensional magneto-optical trap (2D MOT) near the permanent magnets. We then demonstrate the loading of $^{88}$Sr atoms into the tweezer array after multi-stage cooling. With a steady-state 2D MOT being realized with permanent magnets, we find the loading rate of three dimensional blue MOT, located in the separate glass cell,  up to $10^{8}$ s$^{-1}$ at 5 mK. This allows us to trap $2\times10^{6}$ atoms at 2 $\mu$K in a narrow-line red MOT via operating on the $^{1}$S$_{0}$ $\rightarrow$ $^{3}$P$_{1}$ intercombination transition at 689 nm. By exploiting the Sisyphus cooling and pairwise loss processes, single atoms are trapped in 813 nm optical tweezers that are generated by the spatial light modulator (SLM).  We have implemented the Gerchberg-Saxton algorithm
	~\cite{Gerchberg.1972} to equalize the depth of the tweezer trap across the array, achieving a uniformity of over 95$ \%$. Moreover, we explore the Sisyphus cooling at a non-magic wavelength tweezer traps, while characterizing the optical tweezer array through the  release and recapture measurement. This versatile strontium tweezer platform opens up new possibilities for not only for creating Rydberg atom arrays~\cite{cooper2018alkaline,wilson2022trapping,madjarov2020high, norcia2019seconds}, but also for SU(N) fermions~\cite{Taie.2010,Song.2020f2k,He.2020,Sonderhouse.2020,Zhao.2021}.

	The paper is organized as follows. The characterization of the compact vacuum machine, laser systems and optimization for the single-frequency red MOT is detailed in Sec.(\ref{\romannumeral2}). Sec.(\ref{\romannumeral3}) demonstrates the strontium fluorescence in individual programmed optical tweezers loading from the red narrow MOT, and also shows the approaches for alignment of the objective to our vacuum machine. In Sec.(\ref{\romannumeral4}), we investigate the cooling mechanism in 813 nm optical tweezers with linearly polarization under the absence of magnetic field, and have the characterization research of the 813 nm optical tweezers. In the end, Sec.(\ref{\romannumeral5}), we have the summary for our optical-tweezer system in the compact vacuum machine.

	\section{Experimental Setup}
	\label{\romannumeral2}
	
	\subsection{Vacuum system}

\begin{figure}
		\includegraphics[width=3.2in]{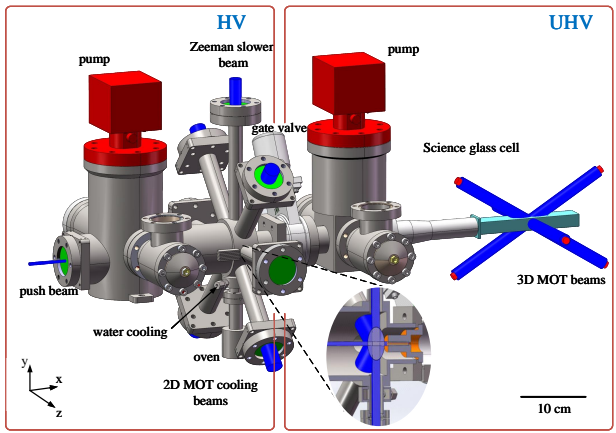}
		\caption{(Color online)
			{Schematic diagram for the vacuum machine and lasers}.
			The vacuum chamber consists of two parts: the HV where strontium is heated from oven and trapped in a 2D MOT; the UHV for further 3D cooling and operation of tweezer arrays. The size of the entire machine is 500$\times$500$\times$200mm in xyz axis and its weight is roughly 50kg.}
		\label{fig1}
	\end{figure}

	Fig.(\ref{fig1}) presents the schematic diagram for the vacuum system and the laser light used to achieve narrow and broad single-frequency MOTs. The entire system, which weighs approximately 50 kg, is mounted on a one-dimensional translation stage to simplify the installation of coils and facilitate vacuum baking. A 30 mm long differential pumping tube with an inner diameter of 2 mm is utilized to maintain balance between the two vacuum chambers. The -x side operates at a high vacuum (UV) level of 10$^{-9}$ Pa, while the x side functions under ultra-high vacuum (UHV) parameters, reaching 10$^{-10}$ Pa when the oven (DN16CF) is not heated. A getter ion pump (SAES Getters, NEXTorr D500) is attached to the main chamber to maximize the pumping speed.
	
	To generate the 2D MOT, the quadrupolar magnetic field is created by four stacks of permanent magnets~\cite{Tiecke2009,Lamporesi2013,Nosske2017,PhysRevA.108.023719} in the y-z plane. Each stack consists of 8 pieces of single N35 neodymium (Nd2Fe14B) magnet (25 mm $\times$ 10 mm $\times$ 3 mm) and the magnetization of 6.6(1) $\times$ 10$^{5}$ A/m. The inset in Fig.(\ref{fig1}) illustrates the 2D MOT configuration with strontium atoms (15g, Sigma-Aldrich Co.,Ltd). To avoid the hot strontium flux coated on the top view port, which can block the transmission of Zeeman slower beams~\cite{PhysRevA.108.023719,li2022bi,PhysRevApplied.19.064011}
	and reduce the atomic flux, we heat up the top sapphire window (located 500 mm away from the oven) at 330\textcelsius. Here, the heating device is a aluminium cylinder with the interlayer being filled with heat band and aluminium silicate thermal insulation. All the view ports can be flexibly connected to a standard 60 mm cage system for mounting optics. A collimated push beam with a 2~mm diameter propagates along the x-axis, producing a steady-state strontium flux. To minimize atom loss during the transport between two chambers, the 2D MOT is situated 2~mm from the entrance of the differential pumping tube which generates the volumetric flow of 3.23$\times$10${^{-2}}$L/s.

	The UV and UHV chambers are separated by an all-metal gate valve (VAT Co., Ltd). In the UHV region, the main science chamber is a customized glass cell (Akatsuki Tech Co.,Ltd), with a size of 150 mm\texttimes17 mm\texttimes27 mm and the wall thickness of 3.5 mm. The glass cell plates are connected by optical bonding (optical contact) method, which can achieve as low as 10${^{-9}}$ Pa. The three retro-reflected MOT beams, with wavelengths of 461 nm and 689 nm, intersect at the center of the glass cell. The two vertical MOT beams intersect at an angle of 120 degrees to provide adequate space for the objective lens, which has a working distance of 15 mm.
	
	To offset the shallow 60-degree angle of the MOT beams relative to the gravity axis, the quadrupole coils (not shown) are molded into an oval shape, with the short axis aligned vertically. These coils are formed by winding a hollow square copper wire that is 5 $\times$ 5 mm in size and features a 3 mm diameter hollow space. This setup enables the generation of a 900 G magnetic field in both reverse and forward directions, using an H-bridge switch system.

	\subsection{Blue MOT}
	
	\begin{figure*}[!htb] 
		\includegraphics[width=7in]{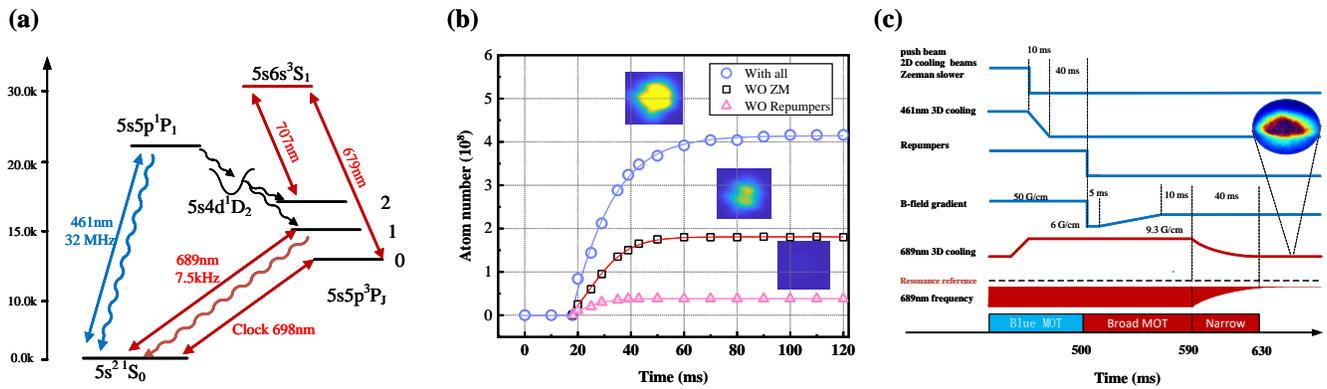}
		\caption{(Color online)
			\textbf{}
			(a) The energy levels for the strontium and the laser system of 461 nm. The blue MOT cooling cycle is compose of a broad cooling transition $^{1}$S$_{0}$ $\rightarrow$ $^{1}$P$_{1}$ at 461 nm, and two repump transitions $^{3}$P$_{2}$ and $^{3}$P$_{0}$ $\rightarrow$ $^{3}$S$_{1}$ at 707 nm and 679 nm respectively to form a closed cooling transition cycle. (b) The loading rate of blue MOT with/without Zeeman slower (ZM) beams and repumper beams.  (c) Time sequence for realizing a narrow red MOT. The inset displays an in-situ image of the compact red MOT.
		}
		\label{fig2}
	\end{figure*} 
	

	We primarily utilize two optical transitions: a broad-line transition at 461 nm and a narrow-line transition at 689 nm. As depicted in Fig.(\ref{fig2}a), the natural line width of the transition between the 5s${^2}{^1}$S${_0}$ and 5s5p${^1}$P${_1}$ states is 32 MHz at 461 nm. This allows for the initial cooling stage, known as blue MOT. But there is a 1:50000 weak leakage, that the atoms drop down from 5s5p${^1}$P${_1}$ to the intermediate state 5s4d${^1}$D${_2}$, which has several hundred $\mu$s life time. Subsequently distribute to the two channel of 5s5p${^3}$P${_1, _2}$. For the blue cooling transition, the ${^3}$P${_2}$ is a dark state with an order of 10 s lifetime. In order to have a closed cooling cycle, two repumper beams 707 nm and 679 nm to excite the atoms to 5s6s${^3}$S${_1}$, only one leakage channel makes the atoms populated in ${^3}$P${_1}$, finally it drops down to the ground state 5s${^2}{^1}$S${_0}$. A high-power 461nm laser (Second Harmonic Generation, Toptica Co.,Ltd), stabilized by a wavelength meter (WS7-30, HighFinesse) is utilized to cool high-temperature strontium atoms expelled from the oven. As illustrated in Fig.~\ref{fig1}, a pair of 2D MOT beams with the waist of 7~mm forms a 2D MOT. 
	

	A red-detuned push beam transports atoms into the 3D MOT region in the science glass cell. In the science glass cell, a 461~nm broad-line 3D MOT is formed with the repumping beams at 679
	~nm and 707
	~nm (see Fig.~\ref{fig2}a), which are transmitted through the 689 MOT polarization-maintaining fibers. The natural linewidths of the repumping beams are 1.75 MHz and 8.9 MHz, and their power intensity is 4$I_{s}$, 7$I_{s}$ respectively. More details on the laser parameters are given in Table \ref{tab:beam parameters}. In Fig.(\ref{fig2}b), we characterize the loading rate of the steady-state blue 3D MOT (around the temperature of 4.7 mK) with Zeeman and repumper beams, respectively. The atom number can reach a stable level of 4$\times$10$^{8}$ when both Zeeman and repumper beams are present. However, when the Zeeman (rempumper) beam is removed, the atom number decreases to 1/3 (1/8) of its initial value.

	\begin{table}[t]
		\caption{Laser Parameters}
		\centering
		
		\begin{threeparttable}
			\begin{tabular}{p{3cm}p{1.8cm}p{1.8cm}p{1.8cm}}
				\hline \hline 
				Beams          & Detune                      & Waist                          & Intensity  \\ 
				& (/$\Gamma$)\tnote{*}          & /mm                             & (/$I_{s}$)\tnote{**}   \\ 
				\hline
				461 2D              & -1.56                  & 7                              & 2.4  \\
				Zeeman slower       & -7.38                  & 7                              & 0.6  \\
				Push                & 0.3                    & 0.7                            & 0.07 \\
				461 3D Hor          & -1.06                  & 5                              & 0.1 \\
				461 3D Vert         & -1.06                  & 6                              & 0.24 \\
				461 probe           & -0.75                  & 5                              & 0.003 \\
				679  repumper       & $\sim 0$               & 6                              & 2.9   \\
				707  repumper       & $\sim 0$               & 6                              & 15    \\
				689 broad Hor       & -667 to -13            & 5                              & 1101(total) \\
				689 broad Vert      & -667 to -13            & 6                              & 2643(total) \\
				689 narrow Hor      & -26                    & 5                              & 58 \\
				689 narrow Vert     & -26                    & 6                              & 140 \\
				\hline\hline
			\end{tabular}
			\begin{tablenotes}
				\item[*]  $\Gamma$ denotes the natural line width of 461 nm, 679 nm, 689 nm, and 707 nm are 32 MHz, 4 MHz, 0.0075 MHz, and 7 MHz respectively;
				\item[**]  $I_{s}$ denotes the saturation intensity of the 461 nm, 679 nm, 689 nm, and 707 nm transition and is given by 43, 0.59, 0.003 and 2.441  mW/cm$^2$ respectively;  	
				
			\end{tablenotes}
		\end{threeparttable}
		
		\label{tab:beam parameters}

	\end{table}
	
	\subsection{A narrow-line red MOT} 
	
	A narrow-line transition at 689~nm enables cool the atoms to a temperature as low as 0.18 $\mu$K in the Doppler cooling limit. In order to narrow the laser to the natural linewidth ($\sim$10 kHz), we utilize a fast analog linewidth control (Toptica, FALC pro) to lock the laser to a commercially available ultra-stable cavity with a finesse of 20,000. To enhance atom loading into the narrow-line 689~nm MOT from the 3D blue MOT, first, we ramp down the blue MOT intensity to zero to obtain a lower temperature, but do not turn off the repump lasers to pump the residual atoms back to the ground states with 40 ms; second, we broaden the laser linewidth to 5 MHz by modulating the RF driving signal of the double pass Acousto-Optic Modulator (AOM) at a frequency of 40 kHz~\cite{qiao2019ultrastable} which generates 125 distinct "comb" peaks spanning the entire range of modulation.
	
	The detailed experimental sequence for realizing a narrow MOT is shown in Fig.(\ref{fig2}c). After the blue MOT is established, we switch off the 2D MOT lasers,  the push beam, 2D cooling beam, and Zeeman slower laser simultaneously. Then, within 10 ms, the intensity of the 3D blue MOT cooling beam is adiabatically ramp down to zero. After additional 40 ms, the 707 nm and 679 nm repumping lasers are immediately turned off.

	The transitions for $^{1}$S$_{1}$ $\rightarrow$ $^{1}$P$_{1}$ and $^{3}$P$_{1}$ require significantly different magnetic field gradients. To this end, we employ a commercially available high-bandwidth PID (Proportion Integration Differentiation) device (Newport LB2005) that takes only 100 $\mu$s to rapidly decrease the magnetic field from 50 G/cm to 6 G/cm. The 689nm cooling light has been continuously enabled, accompanied by a frequency modulation within a range of 5 MHz. 
	
	With 5 ms loading into the red MOT, the magnetic field gradient begins to ramp up to 9.3 G/cm for compressing the atom cloud. The entire process takes about 90 ms, resulting in the atom temperature of 150 $\mu$K. To achieve lower temperature, the modulation frequency gradually decreases to zero following an exponential damping. The MOT beam intensity also ramps down with a exponential function. Achieving a narrow red MOT with 200 kHz red detuning from the resonance of $^{1}$S$_{0}$ and $^{3}$P$_{1}$, we obtain $2\times10^{6}$ cold strontium atoms with the temperature of 2 $\mu$K. This greatly assists us in loading atoms into optical traps, with the specific details to be discussed in the upcoming section.


	\section{Optical tweezers for trapping atoms}
	
	\label{\romannumeral3}
	The creation of tweezer arrays using the spatial
	light modulator (SLM) often faces a trade-off between the quality of reconstruction and the time required for calculations. We have successfully loaded cold strontium atoms into various optical tweezer patterns generated by a SLM integrated with generative neural network~\cite{ren2024creation}. This approach  shortens the process time to control the SLM with minimal time delay, eliminating the need for repeated re-optimization of the hologram for the SLM.
	
\begin{figure}
		\includegraphics[width=3.3in]{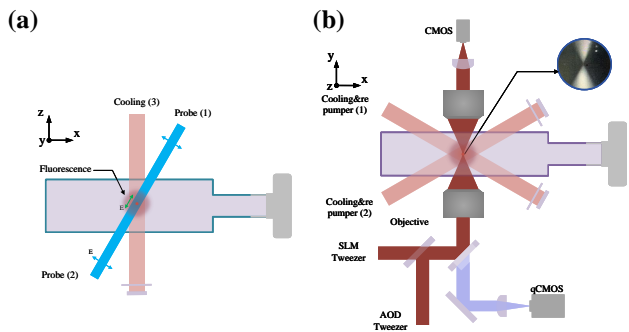}
		\caption{(Color online)
			{Overview diagram for the tweezer laser setup.}
			(a) Top view of the optical layout. The exciting light, intersects at an angle of approximately 10 degrees with the cooling (MOT) beam (3), and is composed of a pair of counter-propagated probe beams with a frequency offset of approximately 0.5 MHz to avoid the optical interference. The power density of the two probe beams can be meticulously fine-tuned independently, guaranteeing an equilibrium of scattering forces on the atoms during fluorescence imaging. The quantum axis is defined by the polarization of the optical tweezer beam which aligns with the propagating direction of the probe beam, the polarization of probe beams are in the zx plane.
			(b) The microscope objective possesses a working distance of 15mm, a focal length of 4mm, and a focal depth of 1.5 $\mu$m. It is securely affixed to a five-axis electric translation stage (not show). Below the objective lens, the optical tweezer laser and imaging fluorescence are combined with a dichroic mirror. At the top, a camera is used to detect the information of optical tweezers. The inset is the atomic fluorescence excited by a probe beam, propagating from the bottom objective lens to the top. The atomic flux travels in the x axis direction from the 2D MOT.}
		\label{fig3}
	\end{figure}

	As shown in Fig.(\ref{fig3}a), the horizontal MOT beam propagates along the z-axis, while also providing cooling during the fluorescence imaging process. The probe beams, intersecting at an angle of approximately 10 degrees with the cooling (MOT) beam (3) in the x-z plane, consists of a pair of counter-propagating beams with a frequency offset of approximately 0.5 MHz to avoid the optical interference. The intensity of probe beam is stablized during fluorescence imaging. In our experimental setup, we choose to establish a zero external magnetic field, as this have the advantage of not only reducing the complexity of calculating energy shift, but also having the capacity to maximize the effectiveness of fluorescence imaging.
	
	The 813~nm optical tweezer beam propagates along the x-axis, and its polarization sets the quantum axis when a linear polarized propagating optical modes with a cross section comparable to the wavelength. Assuming tweezer beam propagate along z axis, the fictitious magnetic fields induced by AC stark shift can be written as\cite{PhysRevA.94.061401} 
	
	\begin{equation}\label{eq:1}
		\begin{aligned}
			\vec{B}_{\rm fic}=-\frac{\alpha^{v}}{\mu_{B}g_{J}J}(-2u_{z}\theta\frac{y}{\omega_{0}}|\varepsilon^{0}_{\rm max}|^{2}e^{-2(z^{2}+y^{2})/\omega^{2}_{0}})
		\end{aligned}
	\end{equation} 

	Here $\alpha_{v}$ is vector polarizability, $\omega_{0}$ is the beam waist radius, $\mu_{B}$ is the is Bohr magneton, $u_{z}$ is the unit vector along z, $\vec\varepsilon$ is the complex unit tweeezer polarization vector, $\varepsilon^{0}_{\rm max}$ is the value $\varepsilon^{0}$ at the origin. The 813 nm optical tweezer laser with polarization aligned in the imaging beam direction, serving as the quantization axis for internal state transitions. The two counter-propagating 461 nm probe lasers propagate horizontally in the radial plane, where the atoms trapped by a high trap frequency.
	
	The AOD (not used in this work) and SLM provide dynamic and static optical tweezers arrays, respectively. At the top, a CMOS camera directly monitor the optical tweezers for real-time calculation of the SLM phase map.
	
	In our experiment, we create tightly focused optical tweezer traps with the beam waist of the 1 $\mu$m range for single atom trapping. As shown in Fig.(\ref{fig3}b), we have chosen a pair of commercial microscope objectives (Mitutoyo, G Plan Apo 50x) with a numerical aperture (NA) of 0.5. Both objective lens is installed on a five-axis electric translation stage (Newport, 8082-M), which has the minimum precision of 30 nm. We delicately manipulate the tilt angle and precise placement of the objective lens so that the fluorescence pathway is parallel to the tweezer beam. To estimate the focal position of the tweezer trap, we use a probe beam from bottom up to excite the fluorescence while activating 2D MOT and push beam.  
	
	\section{Trapping atoms in a tweezer array}
	\label{\romannumeral4}
	
	\subsection{813~nm optical tweezer}
	
	\begin{figure}
		\includegraphics[width=3.5in]{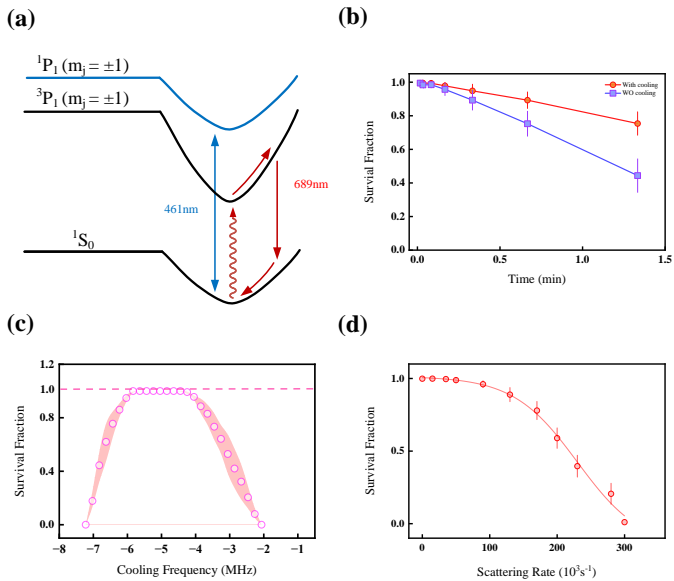}
		\caption{(Color online)
			\textbf{}
			(a). The simplified diagram for broad-line imaging on $^{1}$P$_{1}$ and narrow-line attractive Sisyphus cooling on the $^{3}$P$_{1}$ states. The excited state is characterized by a more profound potential depth than that of the ground state. Atoms initially at the trap of ground state, when excited to the excited state, experience a more steep climb than their original state in the ground state. Following spontaneous emission, the disparity between the energy levels leads to a reduction in the atom's average energy~\cite{covey20192000, urech2022narrow}.			 
			(b). The survival fraction of tweezer array at different trap hold times, with the blue and red curves corresponding to the scenarios with and without Sisyphus cooling, respectively.
			(c). Depicting the relationship between the survival rate of atoms in the array and the cooling frequency detuning from the AC Stark resonance.
			(d). The survival fraction of atoms with the rate of scattered 461 nm imaging light with Sisyphus cooling.
		}
		\label{fig4}
	\end{figure}

	The 813~nm tweezer trap, with a waist of $\sim$700~nm, exhibits multiple intensity maxima with the intensity distribution being a sinc function in the focal plane. To minimize atom loading into subsidiary intensity maxima~\cite{Number2020Niamh,PhysRevA.75.013406}, we first start the optical tweezer with a shallow trap depth of 24 $\mu$K, and subsequently ramp up to 543 $\mu$K. When loading atoms into the optical tweezers, a single atom achieved through a light-assisted collision (LAC), which induces pairwise atom loss using a red-detuned 689 nm light~\cite{cooper2018alkaline,PhysRevX.9.011057,PhysRevX.12.021027}. While the pairwise loss is sensitive to the trap depth~\cite{PhysRevX.8.041054}, we speed up this process to 30 ms by exciting atoms to the $^{1}$P$_{1}$ state using a red detuned 461~nm light.
	
	To have a single-site and single-atom-resolved fluorescence imaging, we collect fluorescence photons with a high-sensitivity qCMOS (Hamamatsu ORCA-Quest C15550-20UP) camera. The frequencies of these two probe laser beams are adjusted to be red detuning 24 MHz from the AC stark shifted resonance in the optical tweezers, and the power intensity (0.001 $I_{s}$) is balanced to each other, ensuring an appropriate scattering rate no more than the cooling rate. During the imaging, we apply Sisyphus cooling to mitigate the heating process. As shown in Fig.(\ref{fig4}a), the excited state $^{3}$P$_{1}$ and  $^{1}$P$_{1}$ potential are deeper than the ground state $^{1}$S$_{0}$, due to the AC stark shift. Atoms initially at the trap of ground state, when excited to the excited state, experience a more steep climb than their original state in the ground state. Following spontaneous emission, the disparity between the energy levels leads to a reduction in the atom's average energy~\cite{covey20192000, urech2022narrow}. 
	
	Fig.(\ref{fig4}b) shows the lifetime achieved using the optimized Sisyphus cooling and probe beams, and the absence of the cooling beam. Without the cooling beam, the lifetime is only around 1 minutes. However, with the application of the cooling beam, it effortlessly extends to 2.5 minutes. With a constant trap depth of $k_{\rm B}$$\times$543 $\mu$K, the high survival fraction is achieved in the cooling beams with red detuning from -1.4 MHz to -1.8 MHz, as illustrated in Fig.(\ref{fig4}c). For more precise measurement, we use release and recapture method to determine a narrower and effective cooling frequency range. In Fig.(\ref{fig4}d), it shows the relation between the scattering rate and the survival fraction, below the 35 kHz, the survival fraction can be reaching to the asymptotic line.

	\begin{figure}[!htb] 
		\includegraphics[width=3.5in]{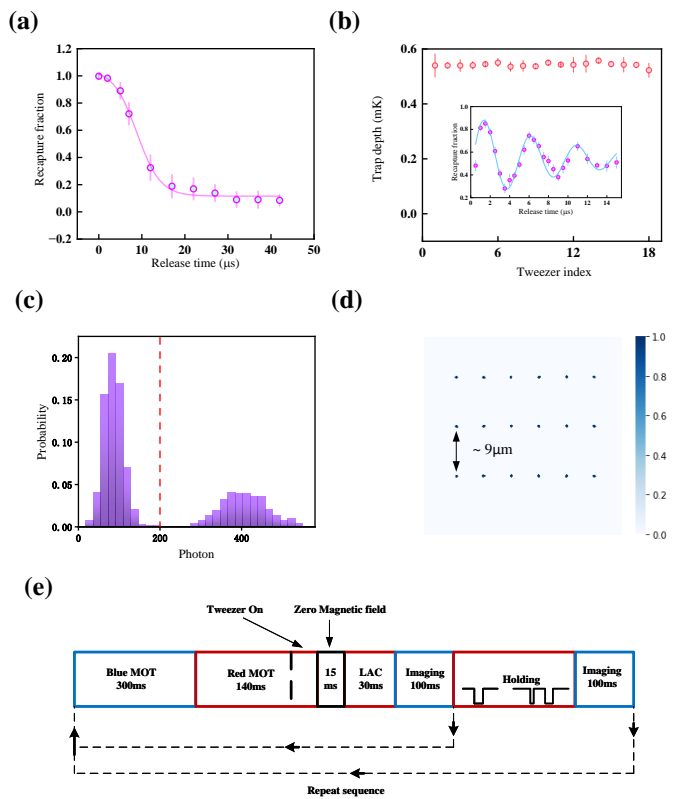}
		\caption{(Color online)
			\textbf{}
			(a). Measured temperature of single atoms trapped in optical tweezers using release and recapture method. Through Monte Carlo simulations, the atoms temperature of the system is approximate 27 $\mu$K.   
			(b). The trap depth of 18 single atoms in tweezers array. It has a standard deviation of around 1.5$\%$ for the tweezer trap depths measured on the atoms transition. The inset shows the measured radial frequency of 103(2) kHz using release and recapture technique.  
			(c). The histogram displays the photon counts of the fluorescence in the region of interest (ROI) of all the single tweezers during the imaging process. Two distinct peaks for vacancy and single atom distribution can be distinctly discerned, which both adhere to a Poisson distribution. The ratio of their integral areas is around 1, indicating a loading efficiency of 0.5. The approximate bin value of the first peak is around 320. The total imaging time is 100ms.
			(d). Averaged single atoms fluorescence of 50 images.
			(e). Experimental sequences for the measurement of histogram, lifetime, fluorescence imaging and temperature of the single atoms, and also the trap frequency. The inset sequences in Holding module are corresponding to the measurement of atom temperature and trap frequency. The repump lasers at 679 nm and 707 nm are switched on during the fluorescence imaging in case of the leakage of $^{1}$D$_{2}$ anti-trapped in 813 nm optical tweezers.
		}
		\label{fig5}
	\end{figure}

	
	

	\subsection{Characterizing tweezers}
	
	To  estimate the temperature of trapped atoms in the tweezer trap, we use release and recapture method. In Fig.(\ref{fig5}a), we vary the release time and measure the survival fraction, from which the temperature is around 27 $\mu$K in comparison to the Monte Carlo result. 
	
	The trap depth of each tweezer affects the Sisyphus cooling effect during the imaging process. To keep the constant trap depth of 3$\times$6 tweezer array, we first detect the tweezers pattern using CMOS camera, and subsequently feedback the tweezer intensity based on the Gerchberg-Saxton algorithm~\cite{PhysRevX.4.021034}. The typical trap depth is precisely determined by the atom loss spectroscopy with the intercombination transition on $^{1}$S$_{0}$ $\rightarrow$ $^{3}$P$_{1}$ at 689 nm. Finally, the trap depth of 18 single atoms has a standard deviation of around 1.5$\%$, as shown in Fig.(\ref{fig5}b). Alternatively, the radial oscillation frequency can be monitored as shown in the inset via release and recapture method. In our experimental sequence, the 813 nm tweezer beam switch twice, the initial release time $\Delta t_{1}$, is to enhance the atoms oscillation amplitude, the release time $\Delta t_{2}$, optimized to have a recapture probability. The atoms have a harmonic oscillation in the trap during the holding, the probability of the recapturing atoms varies according to the oscillation at different positions in the trap. Notably, from the insightful inset presented in Fig.(\ref{fig5}b), the radial trap frequencies is 2$\pi\times$103(2) kHz, while the calculated trap depth and axial trap frequency are 543 $\mu$K and 2$\pi\times$26.9(3) kHz respectively.

	The Fig.(\ref{fig5}c) shows the histogram of background and single atoms fluorescence of 18 tweezers array, it is clear observed the distinguished two peaks, separated by the threshold line of 200. Also the equality of the integrated area suggests that the loading efficiency of single atoms is 50$\%$. In our experiment, a 100 ms imaging time is employed to obtain a distinct signal that surpasses the background noise. Finally, we show the tweezer pattern of 3$\times$6 array with a maximal distance of 9 $\mu m$ in Fig.(\ref{fig5}d), by averaging the fluorescence of 50 images. In Fig.(\ref{fig5}e), the diagram illustrates the experimental timing sequence, which is utilized for fluorescence detection and the measurement of trapping frequency. In the case of the histogram and tweezer array pattern, the sequence is reiterated after first imaging. Conversely, for ascertaining the survival fraction, the atomic array is imaged in a repeated dual-imaging process.
	
	
	
	\section{Summary and Conclusion }
	\label{\romannumeral5}
	
	We present a compact experimental apparatus designed for a strontium Rydberg array. By applying Zeeman slower beam, we efficiently decelerate the atomic beam  ejected from the atomic oven, and produce a 2D MOT. Consequently, we transport atoms into the 3D blue MOT, and ultimately achieve a red MOT with a temperature of 2 $\mu$K and a population of 2$\times10^{6}$ atoms, which facilitates the loading of an optical tweezer array formed. We characterize 813 nm optical tweezer traps with imaging and Sisyphus cooling beams, and quantitatively calibrated experimental parameters by means of measuring and simulating trap frequencies and atomic temperatures through release and recapture method and Monte Carlo simulation. Additionally, we have found that employing a three-dimensional red MOT as the Sisyphus cooling beams is enough and could reduce the complexity of the experimental setup, facilitating daily maintenance. Our newly developed tweezer array setup provides a versatile platform for quantum information processing and quantum simulation with alkaline-earth-like atoms~\cite{he2019recent}.
	
	\vspace{10pt}
	\paragraph*{\bf ACKNOWLEDGMENTS}
	GBJ acknowledges support from the RGC through 16306119, 16302420, 16302821, 16306321, 16306922, 16302123, C6009-20G, N-HKUST636-22, and RFS2122-6S04. KW acknowledges support from the Guangzhou and Nansha District Postdoctoral project. CH acknowledges support from the RGC for RGC Postdoctoral fellowship.

	\bibliography{references.bib}
	
\end{document}